\documentclass[12pt]{article}

\usepackage[utf8]{inputenc}
\usepackage[T1]{fontenc}
\usepackage{physics}
\usepackage{geometry}
\usepackage{svg}

\usepackage[square,sort,comma,numbers]{natbib}

 \usepackage{epstopdf}

\usepackage{amsfonts,amsmath,amsthm,amsxtra}
\usepackage{wrapfig,caption,graphics,graphicx,xcolor,epstopdf}
\usepackage{tabularx, multirow, fancybox, lscape,listings,float,rotating,fancyhdr}
\usepackage{hyperref ,subcaption, cleveref, esvect, physics}
\usepackage{xfrac,nicefrac,textcomp} 
\usepackage[us,12hr]{datetime}
\usepackage{algorithm} 
\usepackage{algpseudocode} 
\usepackage{outlines}
\usepackage{placeins}

\def\begc{\begin{center}}       \def\endc{\end{center}}
\def\bega{\begin{array}}        \def\enda{\end{array}}
\def\eeqn{\begin{equation}}     \def\eeqn{\end{equation}}
\def\begn*{\begin{equation*}}     \def\eeqn*{\end{equation*}}
\def\begea{\begin{eqnarray}}    \def\endea{\end{eqnarray}}
\def\begea*{\begin{eqnarray*}}  \def\endea*{\end{eqnarray*}}

  \def\ss{\subsection}

\newif\ifstylew  \stylewtrue

\stylewfalse   
\newif\ifnotesw \noteswtrue

\begin{document}

\title{\vskip -0.5in
Machine learning pipeline for quantum state estimation with incomplete measurements}

\author{Onur Danaci$^{1}$, Sanjaya Lohani$^{1}$, Brian T. Kirby$^{1,2}$$^{\thanks{brian.t.kirby4.civ@mail.mil}}$, Ryan T. Glasser$^{1}$$^{\thanks{rglasser@tulane.edu}}$\\
{\footnotesize $^{1}$ Tulane University, New Orleans, LA 70118, USA} \\
{\footnotesize $^{2}$ United States Army Research Laboratory, Adelphi, Maryland 20783, USA }
\\
{\footnotesize $^{*}$ brian.t.kirby4.civ@mail.mil} \\
{\footnotesize $^{\dagger}$ rglasser@tulane.edu}
}
\maketitle

\begin{abstract}

Two-qubit systems typically employ 36 projective measurements for high-fidelity tomographic estimation. 
The overcomplete nature of the 36 measurements suggests possible robustness of the estimation procedure to missing measurements.  
In this paper, we explore the resilience of machine-learning-based quantum state estimation techniques to missing measurements by creating a pipeline of stacked machine learning models for imputation, denoising, and state estimation.
When applied to simulated noiseless and noisy projective measurement data for both pure and mixed states, we demonstrate quantum state estimation from partial measurement results that outperforms  previously developed machine-learning-based methods in reconstruction fidelity and several conventional methods in terms of resource scaling.
Notably, our developed model does not require training a separate model for each missing measurement, making it potentially applicable to quantum state estimation of large quantum systems where preprocessing is computationally infeasible due to the exponential scaling of quantum system dimension.

\end{abstract}

\flushbottom
\maketitle
\thispagestyle{empty}

\section{Introduction}

The intersection of classical machine learning (ML) and quantum information science (QIS) has recently become an area of intense investigation \cite{carleo2019machine}.
Applications areas are diverse and include, for example, the representation and classification of many-body quantum states \cite{carleo_constructing_2018}, the verification of quantum devices \cite{lennon2019efficiently}, quantum error correction \cite{nautrup2019optimizing}, quantum control \cite{kalantre_machine_2019}, and quantum state tomography (QST) \cite{lohani2020machine,zimmermann_high-resolution_2011}.  
Early results indicate that ML approaches to processing classical information associated with executing QIS protocols may certain have advantages when compared with standard methods such as improved resource scaling \cite{xu_neural_2018} and resilience to noise \cite{lohani2020machine}.

The estimation of an unknown quantum state with QST is performed using repeated joint measurements on an ensemble of identically prepared quantum systems \cite{altepeter2005photonic, zimmermann_high-resolution_2011}. 
The computational resources required for state estimation alone, after the experimental aspects of QST have been performed, scale poorly even in situations where specific noise models are assumed \cite{smolin_efficient_2012,qi_quantum_2013,hou_full_2016}.
Recently, various ML approaches for quantum state estimation have been proposed \cite{torlai_neural-network_2018,carrasquilla_reconstructing_2019,torlai_latent_2018,xin_local-measurement-based_2018,palmieri_experimental_2019,lohani2019dispersion, lohani2020machine} with some techniques indicating a scaling of $O(d^3)$\cite{xu_neural_2018}.

In principle, two-qubit state tomography is possible from the statistics of only $16$ measurements \cite{vrehavcek2004minimal, zhu2014quantum}. 
However, motivated by results showing that the use of mutually unbiased bases for state estimation minimizes statistical error, two-qubit state tomography is typically performed using $36$ different projective measurement settings \cite{wootters1989optimal,vrehavcek2004minimal}.
Better understanding the reliance of state estimation fidelity on the number of input measurements is an essential question for QST on large quantum systems.  

Enabling an experimentalist to determine how costly avoiding specific measurement settings would be for a given experiment is of significant value.

In general, incomplete QST will not specify a unique state, requiring an additional constraint to decide between physically valid estimations.
One of the most popular approaches is the maximum entropy principle, which finds the state consistent with the measured data with the largest von Neumann entropy \cite{buzek1998quantum}.
Alternatively, variational quantum tomography attempts to find a physically valid state that minimizes the expectation value of the missing projectors \cite{maciel2011variational,gonccalves2013quantum}.
Similarly, methods which jointly maximize the likelihood and the entropy have been explored \cite{teo2012incomplete,teo2011quantum}.
Additionally, under the assumption of high purity, compressed sensing techniques become practical \cite{gross2010quantum,flammia2012quantum}.
Following an entirely different paradigm in our previous work, we showed that training a neural network on simulated tomography data, in principle, results in estimators that still yield high-fidelity quantum state estimation when some measurements are missing  \cite{lohani2020machine}. 
Nevertheless, this problem required training a different model for each combination of missing data, and the necessity of ${36\choose k}$ for $k$ missing measurements makes it computationally unfeasible when moving to large dimensional systems. Here we aim to solve this issue with a limited stack of models and data imputation.

In this paper, we diversify and significantly expand our machine learning models and deploy them within a pipeline for the task of fast and robust quantum state estimation from the partial tomography measurements of high-dimensional quantum systems. Our approach differs from our previous work, where we deployed machine learning techniques for tomographic estimation of two-qubit states \cite{lohani2020machine}, mainly through the decoupling of various functionalities gathered in one single model into different models. First, we trained a one-dimensional convolutional neural network (Conv1D-Regressor), a two dimensional CNN (Conv2D-Regressor), and an extreme gradient boosting regression model (XGB-Regressor) for pure and mixed quantum states using noiseless, simulated measurement data. We trained three regression models for pure and mixed states respectively, with a total of six regression models for the task of quantum state estimation from noiseless projective measurements. Each of these models is sensitive to different types of projective measurements, i.e., predictors or features. Second, we trained a Conv1D based denoising autoencoder (Conv1D-Denoise), which takes noisy data and outputs noiseless data, using both noisy and noiseless simulated measurement data. Third, we trained a Conv1D based classifier to tell whether a sample of projective measurements (36 measurements or features) is noisy or not (Conv1D-isnoise). Fourth, we trained another Conv1D based classifier to tell whether a sample is a pure or mixed state (Conv1D-ispure). Last, we trained a multivariate imputation by a chained equation model based on extreme gradient boosting estimators (XGB-MICE) to predict missing measurements using numerically synthesized experimental data. 

\begin{figure}
    \centering
    \includegraphics[width=0.7\textwidth]{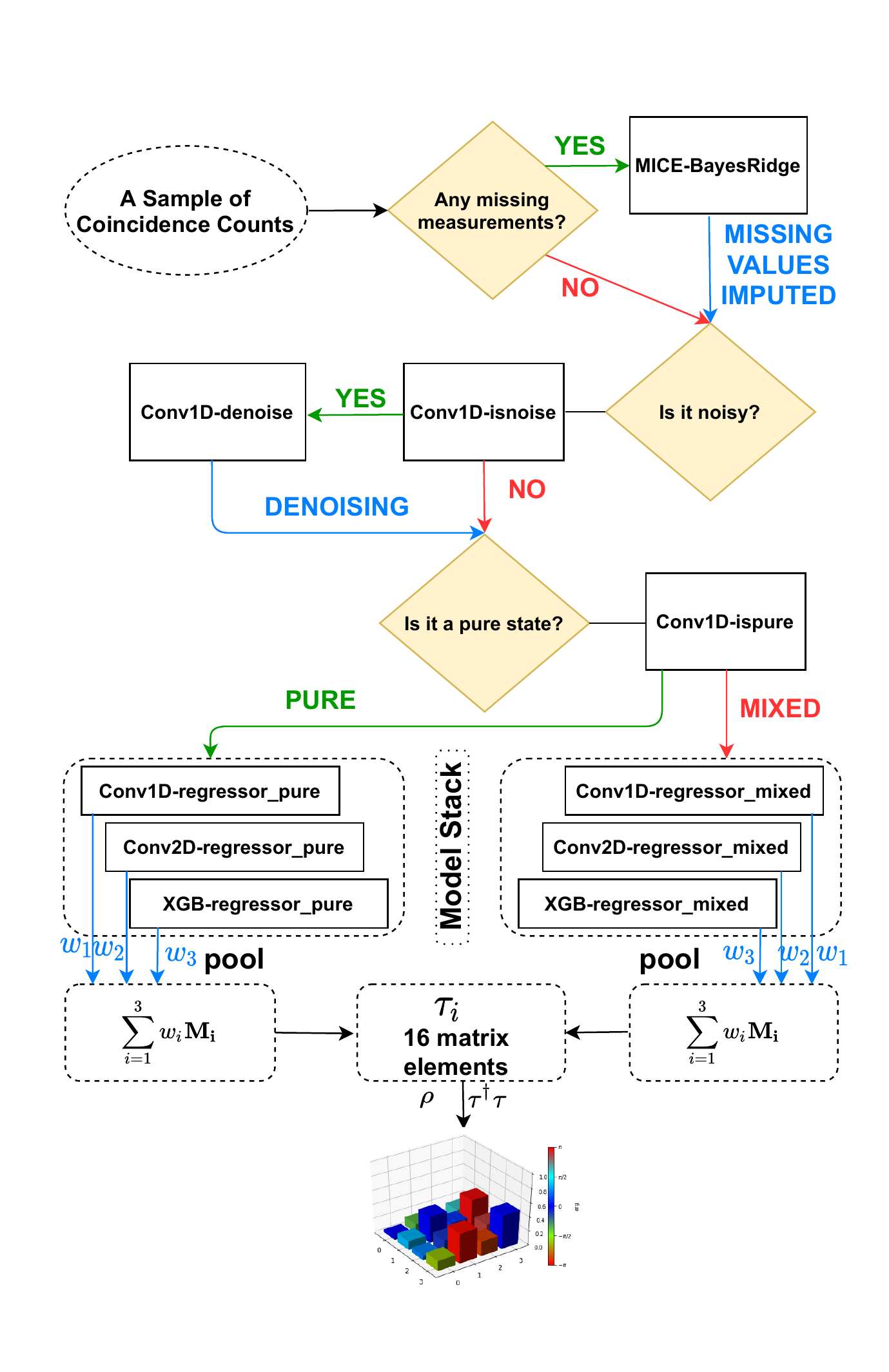}
    \caption{Machine learning pipeline for Quantum State Estimation.}
    \label{fig:stacked-pipeline}
\end{figure}

A schematic of our developed pipeline is shown in Fig. \ref{fig:stacked-pipeline}.  
After training all the models in the pipeline, a sample of a noisy and incomplete set of projective measurements is first complemented by the XGB-MICE model during the testing stage. Next, the complemented sample moves to the Conv1D-isnoise classifier; if it is deemed to be noisy, it moves to Conv1D-Denoise to be denoised, else it moves to Conv1D-ispure. If the Conv1D-ispure model deems the sample a pure state, then the sample proceeds to three fast-QST models trained on pure states (Conv1D, Conv2D, XGB regressors), else it proceeds to their equivalents for mixed states. At the end of the pipeline, three output reconstructed states are obtained from Conv1D, Conv2D, XGB regressors.

Our pipeline is a promising technique for reducing the computational resources required for quantum state estimation.  Having trained on synthesized data, we reconstruct states that outperform our previously developed machine-learning-based approach in reconstruction fidelity and several conventional methods in terms of resource scaling. 
Further, our unique method allows for significant resilience to missing and noisy data without the need to train  ${36\choose k}$ different models for each $k$ missing measurements. 

\section{Methods}

\subsection{Simulating measurement results}

We define an arbitrary set of orthogonal basis states as
\begin{equation}
  |U_{+}\rangle\,=\,\begin{bmatrix}
          1\\
          0\\
         \end{bmatrix}, \quad\, \text{and}\, \quad 
         |U_{-}\rangle\,=\,\begin{bmatrix}
         0\\
         1\\
         \end{bmatrix}.
\label{eqn:HV}
\end{equation}
The orthogonal vectors $\vert U_{+}\rangle$ and $\vert U_{-}\rangle$ can be supplemented with $\vert V_{\pm}\rangle = (|U_{+}\rangle\pm|U_{-}\rangle)/\sqrt{2}$ and $\vert W_{\pm}\rangle = (|U_{+}\rangle\pm i|U_{-}\rangle)/\sqrt{2}$ to form a mutually unbiased bases (MUB).
Further, we notate the tensor product $\vert U_{\pm}\rangle \otimes \vert V_{\pm}\rangle$ as $\vert U_{\pm}V_{\pm}\rangle$, with the other bases combinations treated similarly.
These definitions are general and independent of physical implementation, making them equally applicable to any physical instantiation of a qubit under investigation.
For example, for polarization qubits, the $U_{\pm}$, $V_{\pm}$, and $W_{\pm}$ could be related to the horizontal/vertical, diagonal/anti-diagonal, and right/left circular polarization states, respectively. 

The MUB defined above are over-complete, and a determination of detection probability in each of the six states is sufficient to reconstruct an unknown quantum state \cite{altepeter2005photonic}.
In principle, a state can be reconstructed using a more judiciously chosen set of measurement operators, for a single qubit only four are required \cite{vrehavcek2004minimal}.
However, typical QST systems use the over-complete set defined by the MUB since it minimizes the statistical errors associated with estimating probability distributions from finite samples \cite{wootters1989optimal,vrehavcek2004minimal}.
Extensions of QST to $n$ qubits can be made simply by determining the joint probabilities associated with all $6^n$ measurement combinations on all qubits \cite{altepeter2005photonic}.

We are now ready to summarize the process of QST on an $n$ qubit system; to establish, through repeated measurement, the probabilities associated with all $6^n$ measurement combinations and to then determine the quantum state most consistent with those results.
The inversion of the measurement probabilities into a valid quantum state is the computationally expensive part of QST.
Details about conventional methods for reconstructing the quantum state from measurement statistics using maximum likelihood estimation can be found in \cite{altepeter2005photonic}.
Alternatively, as in our previous work, large amounts of simulated measurement results can be used to train machine learning systems to perform the inversion with generally equivalent performance \cite{lohani2020machine}.

In standard quantum theory the measurement probability $M$ associated with a given projector $P$ is given by $M[P]=\text{Tr}\left(\rho P\right)$.
The aim of QST for $n$ qubits, then, is to simultaneously and consistently invert a set of $6^n$ measurement results $M$ to determine the state $\rho$.
For the two-qubit system we are interested in $36$ joint probabilities must be determined, each associated with a projector from:

\begin{equation}
    P = \begin{bmatrix}
        u_{+} \otimes u_{+}  & u_{+}\otimes u_{-} & u_{-}\otimes u_{-} & u_{-}\otimes u_{+} & u_{-}\otimes w_{+} & u_{-}\otimes w_{-}\\
        u_{+}\otimes w_{-}  & u_{+}\otimes w_{+} & u_{+}\otimes v_{+} & u_{+}\otimes v_{-} & u_{-}\otimes v_{-} & u_{-} \otimes v_{+}\\
        v_{-}\otimes v_{+} & v_{-}\otimes v_{-} & v_{+}\otimes v_{-} & v_{+} \otimes v_{+} & v_{+}\otimes w_{+} & v_{+}\otimes w_{-}\\
        v_{-}\otimes w_{-} & v_{-}\otimes w_{+} &v_{-}\otimes u_{+} &v_{-}\otimes u_{-} & v_{+}\otimes u_{-} & v_{+}\otimes u_{+}\\
        w_{+}\otimes u_{+} & w_{+}\otimes u_{-} & w_{-}\otimes u_{-} & w_{-}\otimes u_{+} & w_{-}\otimes w_{+} & w_{-}\otimes w_{-}\\
        w_{+}\otimes w_{-} & w_{+}\otimes w_{+} & w_{+}\otimes v_{+} & w_{+}\otimes v_{-} & w_{-}\otimes v_{-} & w_{-}\otimes v_{+}\\
        \end{bmatrix},
    \label{eqn:proj}
\end{equation}
where $u_{\pm}\,=\,|U_{\pm}\rangle\langle U_{\pm}|$, $v_{\pm}\,=\,|V_{\pm}\rangle\langle V_{\pm}|$, and $w_{\pm}\,=\,|W_{\pm}\rangle\langle W_{\pm}|$.
For this set of projective measurements the joint detection probability for a given state $\rho$ is given by 
\begin{equation}
    M[6i+j] = Tr(\rho\,P[i,j]); \quad \textrm{for}\,\quad i,j\,=\, 0,\,1,\,2,\,3,\,4,\,5,
    \label{eqn:m}
\end{equation}
Here the matrix has been unrolled into a $1\times 36$ dimensional row vector and the sampling occurs by the row-major ordering of matrices.

Any density matrix $\rho$ reconstructed from a set of measurement results $M$ must be physically valid, that is, it must be a nonnegative definite Hermitian matrix of trace one.  
To ensure this is the case, we do not predict a density matrix $\rho$ directly from measurement results, and instead design our network so that the output (firing of 16 neurons) comprises the elements of the $\tau$-matrix, which can be listed as [$\tau_0$,$\tau_1$,$\tau_2$,$\tau_3$, .. .. .., $\tau_{15}$]. 
These outputs are then rearranged to form a lower triangular matrix as given by, 
\begin{equation}
    \tau = [\tau_0,\tau_1,\tau_2,\tau_3, .. .. .., \tau_{15}] \rightarrow \begin{bmatrix}
                                       \tau_0 & 0& 0& 0\\
                                       \tau_4+i\tau_5 & \tau_1 &0 &0\\
                                       \tau_{10}+i\tau_{11} & \tau_6+i\tau_7 &\tau_2 &0\\
                                       \tau_{14}+i\tau_{15} & \tau_{12}+i\tau_{13} &\tau_8+i\tau_9 &\tau_3\\
    \end{bmatrix},
    \label{eqn:lower_t}
\end{equation}
which always maps to a physically valid density matrix \cite{james_measurement_2001} by
\begin{equation}
    \rho=\frac{\tau^{\dagger}\tau}{\text{Tr}\left(\tau^{\dagger}\tau\right)}.
\end{equation}

For training purposes it is also essential to be able to generate the $\tau$ matrix associated with a given density matrix $\rho$. 
This can be accomplished using the methods of \cite{james_measurement_2001} given by
\begin{equation}
    \tau = \begin{bmatrix}
                 \sqrt{\frac{Det(\rho)}{m_1^{00}}} & 0 & 0 & 0\\
                 \frac{m_1^{01}}{\sqrt{m_1^{00}m_2^{00,11}}} & \sqrt{\frac{m_1^{00}}{m_2^{00,11}}}& 0 & 0\\
                 \frac{m_2^{01,12}}{\sqrt{\rho^33}\sqrt{m_2^{00,11}}}&\frac{m_2^{00,12}}{\sqrt{\rho^{33}}\sqrt{m_2^{00,11}}}&\sqrt{\frac{m_2^{00,11}}{\rho^{33}}}
                 & 0\\
                 \frac{\rho^{30}}{\sqrt{\rho^{33}}} &\frac{\rho^{31}}{\sqrt{\rho^{33}}}&\frac{\rho^{32}}{\sqrt{\rho^{33}}}&\sqrt{\rho^{33}}
                \end{bmatrix}
\label{eqn:tau}
\end{equation}
where $m_1^{ij}$ for $i,j \in \{0,1,2,3\}$, and $m_2^{pq,rs}$ $(p\neq r$ and $q\neq s)$ for $p,q,r,s \in \{0,1,2,3\}$ are the first and second minor of $\rho$, respectively.

Finally, we note that our metric for estimation accuracy between a generated matrix $\rho$ and a target matrix $\rho_0$ is the fidelity, given by
\begin{equation}
    F = \Big|Tr\sqrt{\sqrt{\rho_{pred}}\rho_{targ}\sqrt{\rho_{pred}}}\Big|^{2}.
    \label{eqn:fidelity}
\end{equation}

\subsection{Simulation noisy measurement results}

To simulate noisy measurement results we introduce arbitrary rotations into the projective bases of $P$.
This noise model is experimentally inspired and represents the the difficulty of perfectly determining and aligning measurement bases.
Arbitrary rotations to the measurement bases are applied using the rotational operator $\mathcal{R}$ given by
\begin{equation}
    \mathcal{R}(\vartheta,\varphi,\xi) = \begin{bmatrix}
    e^{i\varphi/2}\cos(\vartheta) & -i\,e^{i\xi}\sin(\vartheta)\\
    -i\,e^{-i\xi}\sin(\vartheta) & e^{-i\varphi/2}\cos(\vartheta),
    \end{bmatrix}
    \label{eqn:rot}
\end{equation}
Where $I$ is the identity matrix.
The variables $\vartheta,\varphi,\xi$ are sampled from the normal distribution with zero mean and $\sigma^2$ variance. Application of this noise model to $M$ results in 
\begin{equation}
    M_{noise}[6i+j]=\text{Tr}(\rho\,(I\otimes \mathcal{R})P[i,j](I\otimes \mathcal{R}^\dagger))
    \label{eqn:rot2}
\end{equation}
In simulations throughout this paper the elements of $\mathcal{R}$ are treated as random variables and are sampled individually for each element $i$ and $j$ of $M$.
Therefore, as in a real QST experiment, the alignment of bases could be variable as the $36$ measurements in $M$ are cycled through.

\subsection{Generating random two-qubit states}

So that our QST system is applicable to all possible input states we train and test it against both pure and mixed states generated at random. 
Random pure states of two-qubits are created by  generating Haar random $4\times4$ unitary matrices and taking the first column as the coefficients of the state.
Specifically, given a Haar random unitary $\mathcal{A}$ the accomponying pure state is 
\begin{equation}
    |\psi\rangle = \mathcal{A}_{00}|U_{+}U_{+}\rangle + \mathcal{A}_{10}|U_{+}U_{-}\rangle + \mathcal{A}_{20}|U_{-}U_{+}\rangle + \mathcal{A}_{30}|U_{-}U_{-}\rangle,
    \label{eqn:psi}
\end{equation}
where $\mathcal{A}_{ij}$ represents the $i^{th}$ row and $j^{th}$ column of $\mathcal{A}$.
Note that we add a tiny perturbation term $\epsilon\,(1\times10^{-7})$ to the simulated pure states as $\rho_{pure}=(1-\epsilon)\vert\psi\rangle\langle\psi\vert+\frac{\epsilon}{4}I$ to avoid the possible convergent issue under Cholesky decomposition \cite{higham_analysis_1990}.

Random mixed states are generated from the Ginibre ensemble \cite{forrester_eigenvalue_2007} given by
\begin{equation}
    G = N\big(0,1,[4,4]\big)\,+\,i\,N\big(0,1,[4,4]\big),
    \label{eqn:gini}
\end{equation}
where $N\big(0,1,[4,4]\big)$ represents the random normal distribution of size of $4\times 4$ with zero mean and unity variance. 
The random mixed state is extracted from this ensemble using
\begin{equation}
    \rho_{mix} = \frac{GG^\dagger}{\text{Tr}(GG^\dagger)},
    \label{eqn:mix}
\end{equation}
where $\text{Tr}$ represents the matrix trace.

\subsection{Generating training sets}

In order to train the models on our state estimation pipeline we start by randomly generating 1 million pure states, 1 million mixed states, and their corresponding tomography measurements. This yields design matrices, $X_{\textrm{noiseless\textunderscore pure}}$, of size $1M\times 36$ for projective measurements from noiseless pure states, $X_{\textrm{noiseless\textunderscore mixed}}$ of size $1M\times 36$ for measurements from noiseless mixed states,  and regression target matrices $Y_{\textrm{noiseless\textunderscore pure}}$ of size $1M\times16$ of $\tau$ elements for pure states, and $Y_{\textrm{noiseless\textunderscore mixed}}$ of size $1M\times16$ for mixed states. We split each of them into training, validation (hold-out), and test sets with ratios of $90\%$ (900K), $5\%$ (50K), and $5\%$ (50K), respectively.

Next, we generate noisy measurements. For that purpose, we generate 400 sets of noisy measurements per density matrix we created. The random rotation angles $\vartheta,\varphi,\xi$ for the unitary matrix given in equation \ref{eqn:rot} are randomly sampled to generate random matrices to create measurements using equation \ref{eqn:rot2}. For each of these sets of 400 measurements (400 random matrices) the angles $\vartheta,\varphi,\xi$ are sampled from the following distributions of mean zero, and standard deviation of $\sigma$; 100 of them from the normal distribution $\mathcal{N}(0,\sigma)$, 100 of them from the Laplace distribution $L(0,\sigma)$, 50 of them from brown noise $Br(0,\sigma)$, 50 of them from blue noise $Bl(0,\sigma)$, and 50 of them from pink noise $Pink(0,\sigma)$. Normal and Laplacian noise are sampled using \textit{Numpy} \cite{oliphant2006guide}, while colored noises are sampled by modifying the \textit{colorednoise} package \cite{colorednoise} following the recipe given in \cite{timmer1995generating}. Next, we create another 9000 pure states and 9000 mixed states, separating each of them into three sets of 3000, then splitting these sets of 3000s into separate training (90$\%$, 2700), validation (5$\%$, 150), and test sets (5$\%$, 150). From each of these sets of 3000 states, we create 1.2M noisy measurements through 400 random unitary rotations such that we get 1080K training, 60K validation, and 60K test measurements per set of 3000. We sample random angles with different standard deviations for each of these in 3 sets of 3000 by standard deviations of $\pi/24$, $\pi/12$, $\pi/6$ such that we generate 3.6 million noisy pure state measurements, and we do the same to the three sets of 3000s for the mixed states to obtain 3.6 million noisy mixed state measurements. The respective training-validation-test sets of each get measurements from different quantum states to prevent over-fitting. e.g., the set of mixed states with noise $\pi/6$ has $1.2M$ measurements is split into subsets of 1080K on training, 60K on validation, and 60K on test sets, but each of these subsets are the noisy measurements of different quantum states. Each of these noisy measurement sets comprises two design matrices $X_{allnoise}$ and $X_{nonoise}$. The former has 400 rows of noisy measurement samples of size $1\times 36$ vertically stacked per density matrix, while the latter has 400 copies of the noiseless measurement stacked per density matrix such that their dimensions match. Also, for the $\tau$ matrix elements to be estimated, each set has the target variable matrix $Y_{nonoise}$. For regression, there are 400 copies of $1\times 16$ $\tau$ samples stacked per density matrix. At the end we have generated 7.2M noisy measurements; 3.6M from pure states (1.2M from $\pi/24$,1.2M from $\pi/12$,1.2M from $\pi/6$), and 3.6 from mixed states (1.2M from $\pi/24$,1.2M from $\pi/12$,1.2M from $\pi/6$).  These correspond to the following three matrices: $X_{allnoise}$ of size $7.2M\times 36$ with all original rows, $X_{nonoise}$ of size $7.2M\times 36$ with 18K original rows (due to 9K pure and 9K mixed), and $Y_{nonoise}$ of size $7.2M\times 16$ with 18K original rows.

Finally, we generate matrices for our classifier, which detects whether a state is pure is mixed.  By vertically stacking one million measurements from pure noiseless states with 3.6 million from pure noisy states, and stacking it with one million from noiseless mixed states and 3.6 million from noisy mixed states, we create the matrix $X_{ispure}$, and correspondingly create the target vector $\vb{t}_{ispure}$ that has the first 4.6 million rows as zeros, and next 4.6 million rows as ones.  Similarly, we vertically stack 1 million measurements from the pure noiseless stack with 1 million from the noiseless mixed states and stack them with 7.2 million from the noisy states to create the matrix $X_{isnoise}$. We generate a target label vector $t_{isnoise}$ that has its first 2 million entries for noiseless states as zero and the next 7.2 million entries as ones.

\ss{Imputation with MICE-BR}

Imputation is a statistical technique to infer missing values of input predictors from the rest of the data on hand. The single imputation using regression (Buck method) first fills the missing values with the sample mean of the given predictor, then fits a linear regression model for each predictor given predictors are highly correlated \cite{buck1960method}. 
The overcomplete nature of state tomography with mutually unbiased bases, totaling 36 measurement probabilities for two-qubits, are strongly correlated for a wide array of input states.
Therefore, linear regression models for each of 36 measurement probabilities using other values can be fitted to predict missing tomography measurements \cite{wang_robust_2009}. Despite the circular dependence, i.e., fitting a model to predict $x_a$ using $x_b$, then fitting another for vice versa, this single imputation technique works remarkably well for such correlated variables \cite{little2019statistical}. In our case, it corresponds to fitting simple linear regression models (design matrix composed only of predictors as features) with the zeroth measurement, $x_0$, as the target variable while using the other $35$ inputs as features, then fitting a model with $x_1$ as targets while using the rest of $35$ measurements as features, with the pattern continuing. With fitted probabilistic linear models, stochastic regression imputation can sample imputation values from the imputation model's normal predictive distribution (e.g., a Gaussian on stochastic regression's predicted mean and variance) fit on features \cite{little2019statistical}. 

We employ a technique known as Multiple Imputation by Chained Equations (MICE), which runs multiple stochastic imputation models with different random seeds that can be used to sample and pool multiple values \cite{buuren2010mice}.
A single imputation algorithm \textit{IterativeImputer} interatews over multiple imputations in a round-robin fashion \cite{buuren2010mice} yet still converging to a single value, making it easily corrupted by noise \cite{little2019statistical}. 
By pooling values obtained from different random seeds, MICE is more robust to noise.

We fit four different Bayesian Ridge (BR) models to the concatenated training sets of noiseless pure ($900K\times36$) and mixed ($900K\times 36$) state measurements (1.8M) with the parameters $\alpha_1=\alpha_2=\lambda_1=\lambda_2=10^{-6}$ under the \textit{IterativeImputer} wrapper.  The BR regressions we fit to predict measurements learn how to regularize the weight of other measurements from the data itself, preventing overfitting (e.g., if measurement 35 is not affecting measurement 0, it will have a smaller weight while regressing for 0). Then, during the \textit{transform} stage we use $t=15$ iterations, and four random seeds on the individual iterative imputations. 

In order to assess training performances of algorithms that impute missing at random (MAR) \cite{rubin1976inference} type data, Monte Carlo (MC) methods can be used to simulate the imposition of missing data into a complete set. As the imposition of MAR resembles a simple probit type of regression (i.e., logistic regression of whether the binary choice of a particular value is missing with a probability), missingness corresponds to  random sampling from the data with respect to a probability density function; i.e., using random masks to turn off values \cite{lang2014supermatrix}. The value to be estimated, such as the mean of MSE between imputed and original data in our case but the value of Monte Carlo integrals in general, can be obtained through random masks shaped by low discrepancy sampling (quasi-random numbers) such as Halton sequences \cite{halton1960efficiency}. 

\begin{figure}[t]
    \centering
    \includegraphics[width=0.75\textwidth]{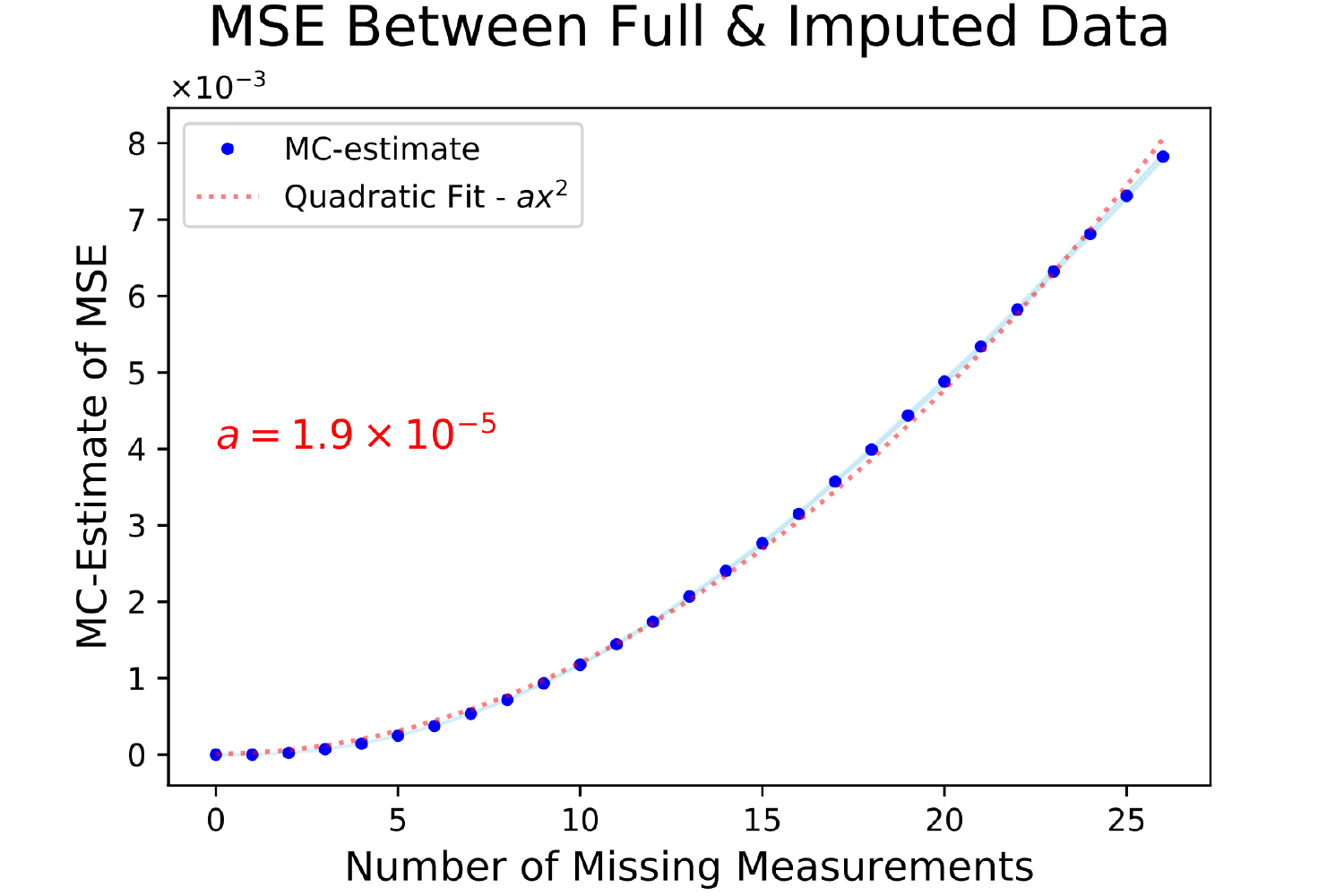}
    \caption{Noise induced by missing values imputation.}
    \label{fig:miss-mse}
\end{figure}

Our quasi-Monte Carlo simulations of the MICE performance involves generating masks that randomly turn off a given number of measurements using Halton sequences. For example, to simulate the case of a single measurement gone missing, we may randomly turn the 20th, 12th, 35th, 3rd (and so on) measurements for the first fourth tomography measurement samples (and so on) into \textit{NaN} (not a number), then impute them using MICE. We take 10K noiseless samples of measurements, create 50 masks of size $10K\times36$ to emulate one measurement missing, then 100 to emulate two, etc., increasing by 50 until reaching 500 masks (turning off ten measurements) and using 500 masks to take MC estimates for any number of missing measurements higher than 10. 
We find that there is almost no noise up to 5 missing measurements, then the MSE quadratically increases as demonstrated by the quadratic polynomial with coefficient $a=1.9\times10^{-5}$ (without intercept or linear term) we fit in fig. \ref{fig:miss-mse}. However, as seen in Fig. \ref{fig:miss-mse}, the more measurements are missing, the more statistical noise is introduced to the recovered measurements.

Having imputed missing measurements using low-level features of MICE-BR, though the contribution weight of each measurement on imputing another is learned from the data, we need more sophisticated algorithms for state estimation. Quantum state estimation, formulated as a regression problem, requires high-level features or high-level predictor segmentation. For that, we turn to the neural network type algorithms, then boosting trees.

\subsection{Convolutional Neural Networks and Auto-Encoders}

\begin{figure}[h!]
    \centering
    \includegraphics[width=\textwidth]{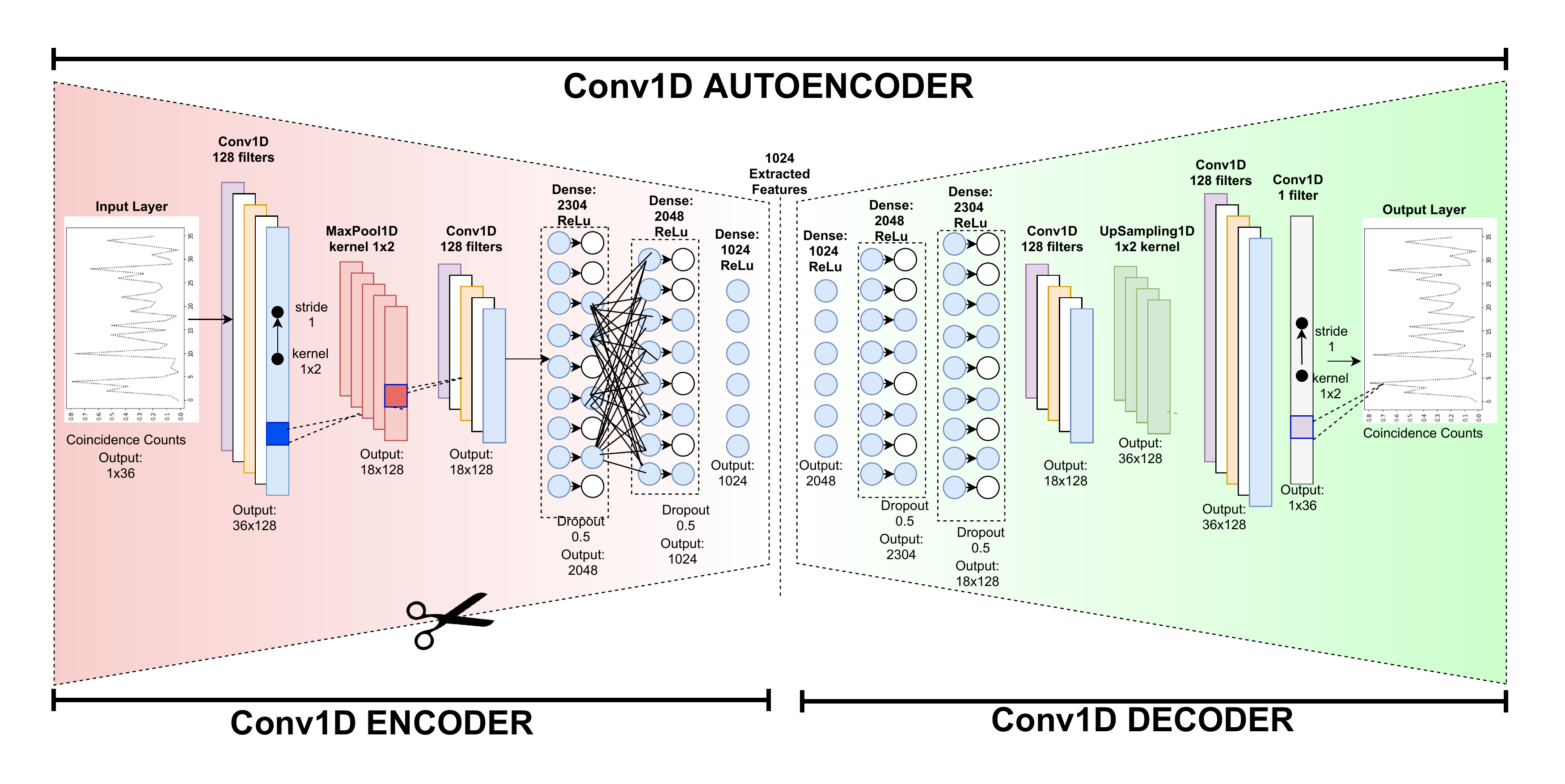}
    \caption{Our 1D convolutional auto-encoder design to be used for denoising and feature extraction for both regression and classification. The scissor icon is to denote the encoder portion is frozen, and cloned to be used in different neural networks.  Our 2D convolutional auto-encoder works in an analogous fashion.}
    \label{fig:conv1d-ae}
\end{figure}

\begin{figure}
    \centering
    \centering
    \includegraphics[width=0.8\textwidth]{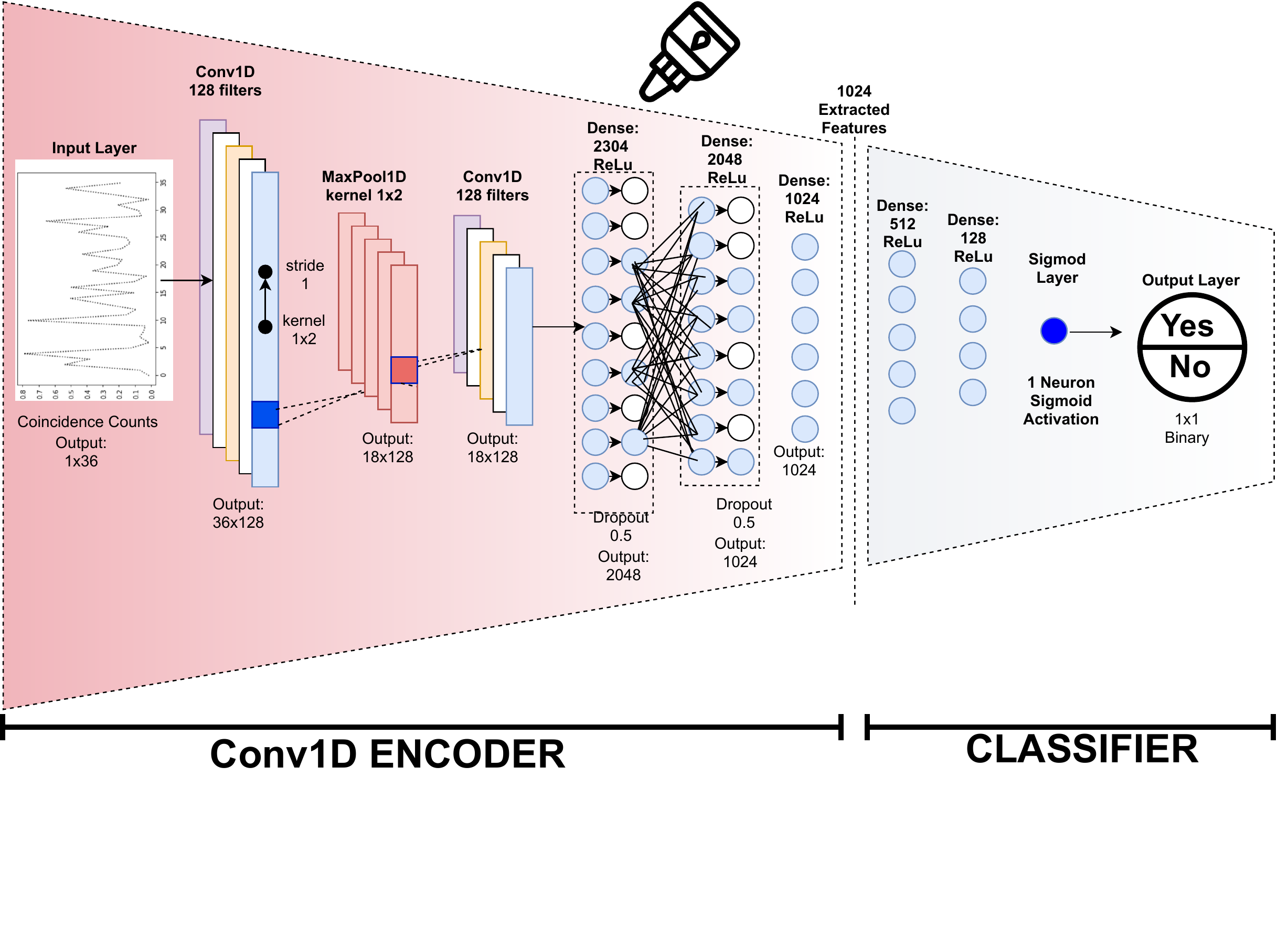}
    \caption{Our NN design for classification.}
    \label{fig:conv1d-clf}
\end{figure}

We train five different unsupervised auto-encoder models on Keras with Tensorflow backend \cite{chollet2015keras}. We use a batch size of 256 and shuffle training data at each epoch for all of our models. First, by supplying a 900K noiseless pure state measurement training set as both the input and the output, while validating on 50K (validation set of noiseless pure states) again as both input and output, we train two auto-encoders that use Conv1D-autoencoder and Conv2D-autoencoder architecture using the MSE loss function.  For clarity we include a visualization of the Conv1D-autoencoder in Fig \ref{fig:conv1d-ae}.  The Conv2d-autoencoder is not pictured, but functions analogously.
We train both of these AEs using one epoch of Adam optimizer with learning rate $\eta=0.003$, two epochs of Adam optimizer with learning rate $\eta=0.001$, one epoch of standard gradient descent (SGD) with learning rate $\eta=0.001$, batch size of 256, and executing early stopping when the validation set MSE reaches $4\times10^{-4}$. As a result we obtain the intermediate models \textit{Conv1D-AE-noiseless\textunderscore pure} and \textit{Conv2D-AE-noiseless\textunderscore pure}. Then, we take two clones of \textit{Conv1D-AE-noiseless\textunderscore pure} and one clone of \textit{Conv2D-AE-noiseless\textunderscore pure}. The clone of \textit{Conv2D-AE-noiseless\textunderscore pure}, and the first clone of \textit{Conv1D-AE-noiseless\textunderscore pure} are re-trained (on top of their previous training as weight initialization) using a 900K noiseless mixed state measurement training set as both the input and the output, while validating on their 50K validation set as the input and output, as the intermediate models \textit{Conv2D-AE-noiseless\textunderscore mixed} and \textit{Conv1D-AE-noiseless\textunderscore mixed}. These two models are both trained using one epoch of Adam optimizer with learning rate $\eta=0.001$, one epoch of SGD with learning rate $\eta=0.001$,  and early stopping set for a validation MSE of $4\times10^{-4}$. The second clone of \textit{Conv1D-AE-noiseless\textunderscore pure} is also trained further by adding noiseless mixed state data on top of the previous data as an intermediate model \textit{Conv1D-AE-noiseless\textunderscore both}. Concatenating 900K from training sets and 50K from validation sets of the noiseless pure and mixed state measurements \textit{Conv1D-AE-noiseless\textunderscore both} model takes these 1.8M measurements as the training data (both as input and output) while validating on the 100K. \textit{Conv1D-AE-noiseless\textunderscore both} is then trained with the exact same epochs and learning rates as the \textit{Conv2D-AE-noiseless\textunderscore mixed}.

\begin{figure}
    \centering
    \includegraphics[width=0.85\textwidth]{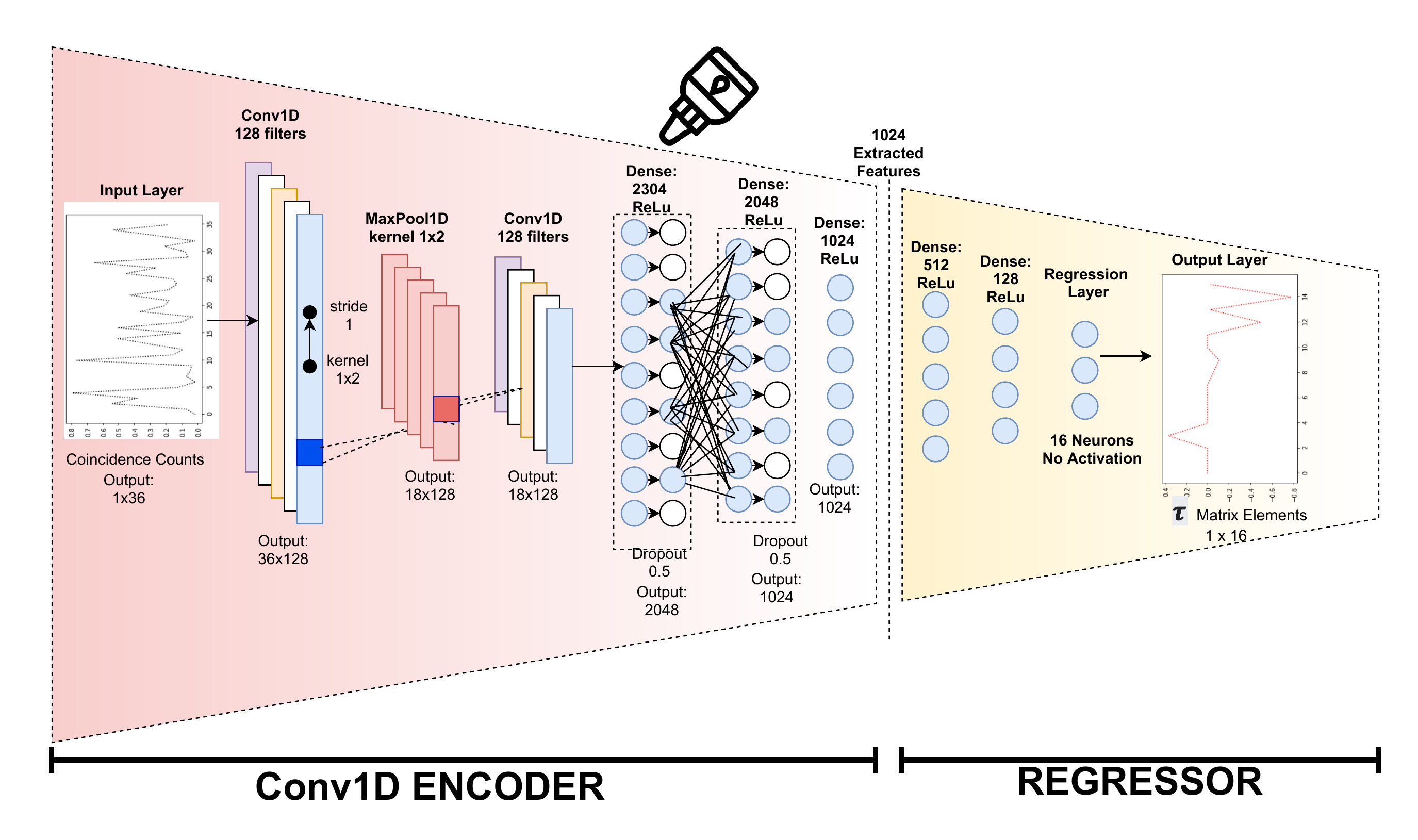}
    \caption{Conv1D-regressor design.}
    \label{fig:Conv1d-regr}
\end{figure}
 
 Second, we transfer the encoder portion of the \textit{Conv2D-AE-noiseless\textunderscore both}, use its extracted latent features and freeze its weights to a NN that uses the Conv1D-Classifier architecture given in Fig. \ref{fig:conv1d-clf} to a model we named \textit{Conv1D-isnoise} in the flow chart given in Fig. \ref{fig:stacked-pipeline}. The classifier portion to be trained is initialized randomly, while the initial encoder layers transferred are not trained. We also clone this model into another classifier named \textit{Conv1D-ispure}. \textit{Conv1D-isnoise} is trained using the training set of the matrix $X_{isnoise}$ given the above as the input, and the training set of the vector labeled $t_{isnoise}$ as the output features while validating on their respective validation sets. Similarly, \textit{Conv1D-ispure} is trained using the training set of the matrix $X_{ispure}$ and the training set of the vector labeled $t_{ispure}$ as the output features while validating on their respective validation set. We use \textit{binary cross-entropy} as the loss function. Due to unsupervised pre-training using auto-encoders, both \textit{Conv1D-isnoise} and \textit{Conv1D-ispure} classifiers obtains a binary prediction performance of $0.99999$ F-score \cite{sasaki2007truth} and AUC \cite{fawcett2006introduction} for the training, validation, and test sets in just one epoch of training with Adam optimizer and learning rate $\eta=0.001$. The training performance details for these classification models are given in the table below.
 
 \begin{center}
 \begin{tabular}{||c c c c||} 
 \hline
 \; & \textrm{Train F-Score} & \textrm{Validation F-Score} & \textrm{Test AUC} \\ [0.5ex] 
 \hline\hline
 \textit{Conv1D-isnoise} & 0.9999 & 0.9999 & 0.9999 \\ 
 \hline
\textit{Conv1D-ispure} & 0.9999 & 0.9999 & 0.9999  \\ [1ex] 
 \hline
\end{tabular}
\label{tab:class_tab}
\end{center}

 Third, we take another clone of pre-trained \textit{Conv1D-AE-noiseless\textunderscore both} to train a supervised denoising auto-encoder model given in Fig. \ref{fig:stacked-pipeline} as  \textit{Conv1D-denoise}. Both in the training and validation stages, \textit{Conv1D-denoise} takes 400 noisy measurements per density matrix as the input and tries to map it to the 400 copies of the original, noiseless measurements from the said density matrix. We use the training set of $X_{allnoise}$ given above as the input and $X_{nonoise}$ as the output features while validating on their respective validation sets. After training for five epochs using Adam optimizer with learning rate $\eta=0.001$, and two epochs with SGD with learning rate $\eta=0.001$, we halt the training when the early stopping criteria of training and validation set MSE of $10^{-3}$ is reached. Bench-marking on the test set, we also observe a testing set denoising performance of $10^{-3}$ MSE.

 Fourth, the encoder portions of \textit{Conv1D-AE-noiseless\textunderscore pure} and \textit{Conv2D-AE-noiseless\textunderscore pure}, and \textit{Conv1D-AE-noiseless\textunderscore mixed} and \textit{Conv2D-AE-noiseless\textunderscore mixed} are transferred to the \textit{Conv1D-regressor\textunderscore pure}, \textit{Conv2D-regressor\textunderscore pure}, \textit{Conv1D-regressor\textunderscore mixed}, \textit{Conv2D-regressor\textunderscore mixed} models, with the one dimensioanl designs shown in Fig. \ref{fig:Conv1d-regr}, with analogous designs for the two dimensional case. Again only the 1024 latent features extracted by the pre-trained AEs encoder portion are used, without further training the transferred encoder portion for all the four models. These encoder portions are horizontally stacked with the regressor NN portions to be trained. \textit{Conv1D-regressor\textunderscore pure} and \textit{Conv2D-AE-noiseless\textunderscore pure} are trained using the training set of $X_{\textrm{noiseless\textunderscore pure}}$ as the input predictors, and $Y_{\textrm{noiseless\textunderscore pure}}$ as the output regression targets, while validating on their respective validation sets. Similarly we train \textit{Conv1D-regressor\textunderscore mixed} and \textit{Conv2D-AE-noiseless\textunderscore mixed} using the training set of $X_{\textrm{noiseless\textunderscore mixed}}$ as the input predictors, and $Y_{\textrm{noiseless\textunderscore mixed}}$ as the output regression targets, while validating on their respective validation sets. We train these four models using three epochs of Adam optimizer with learning rate $\eta=0.001$, and two epochs of SGD with learning rate $\eta=0.001$.  Training of those that meet the early stopping criterion of validation set with MSE $4\times10^{-4}$ are halted before the end of epochs. By computing their quantum fidelities, we found that Conv1D based models have an average fidelity of $1.0$ and a standard deviation of $0.0$ in their respective test sets, and Conv2D ones are also nearly unity. The training performance details for these regression models are given in the table below.
 
  \begin{center}
 \begin{tabular}{||c c c c c||} 
 \hline
 \; & \textrm{Train MSE} & \textrm{Validation MSE} & \textrm{Validation Fidelity} & \textrm{Test Fidelity}\\ [0.5ex] 
 \hline\hline
 \textit{Conv1D-regressor\textunderscore pure}  & $5\times10^{-4}$ & $4\times10^{-4}$ & 1.0 & 1.0 \\ 
 \hline
  \textit{Conv2D-regressor\textunderscore pure}  & $7\times10^{-4}$ & $6\times10^{-4}$ & 0.995 & 0.995 \\ 
 \hline
  \textit{Conv1D-regressor\textunderscore mixed}  & $5\times10^{-4}$ & $4\times10^{-4}$ & 1.0 & 1.0 \\ 
 \hline
  \textit{Conv2D-regressor\textunderscore mixed}  & $8\times10^{-4}$ & $7\times10^{-4}$ & 0.997& 0.997 \\ [1ex] 
 \hline
\end{tabular}
\label{tab:conv_regress_tab}
\end{center}

\ss{XGB-Regression}

\begin{figure}[t]
    \centering
    \includegraphics[width=\textwidth]{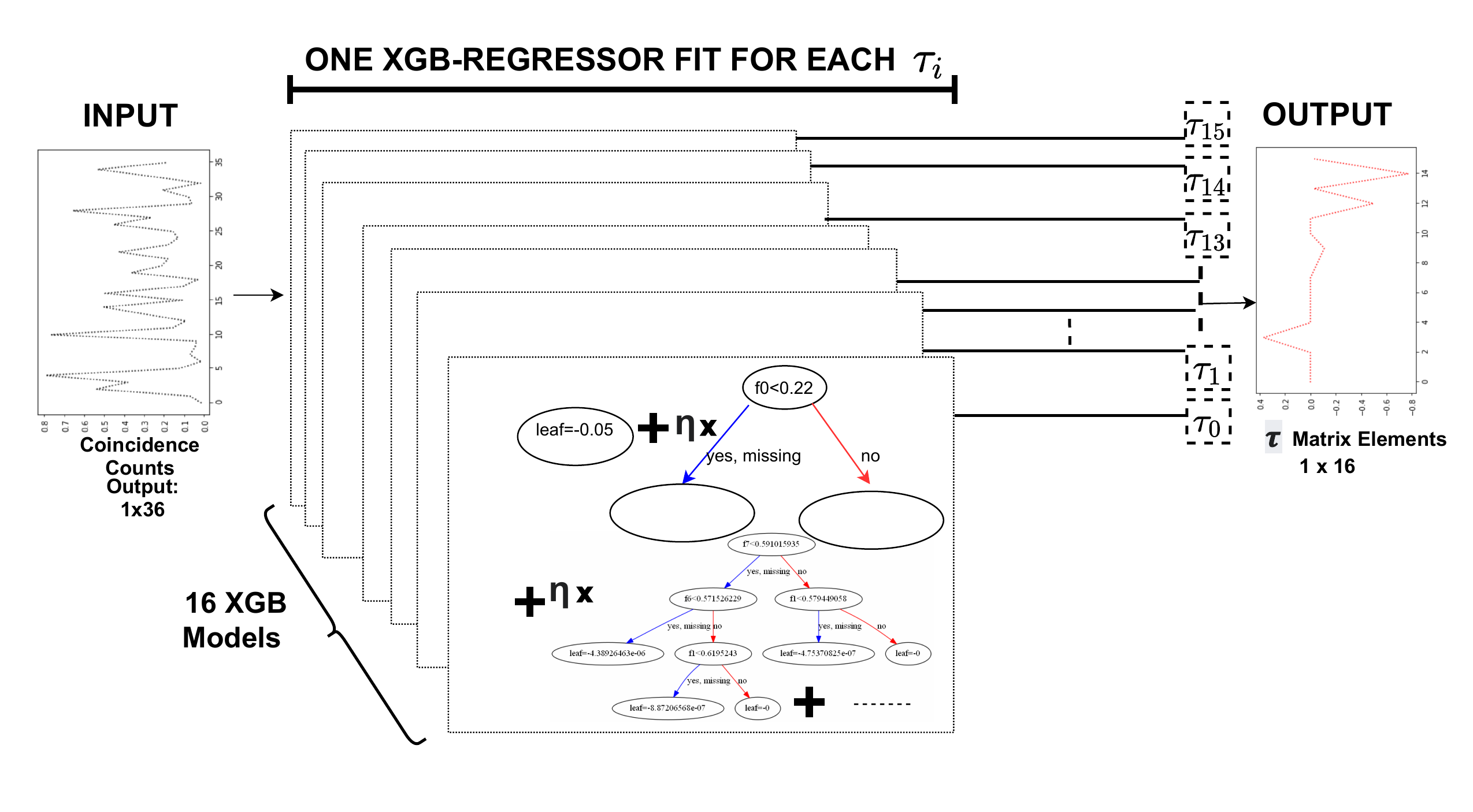}
    \caption{Our multiple target XGB regression design uses 16 separate XGB regression models fit to each of 16 $\tau$ matrix elements, using 36 tomography measurements.}
    \label{fig:xgbmultiple}
\end{figure}

Here we employ the state-of-the-art for tree based algorithms, the Extreme Gradient Boosting Algorithm (XGB) \cite{chen2015xgboost}. 
We use two models, one for noiseless pure states \textit{XGB-regressor\textunderscore pure}, and one for noiseless mixed states \textit{XGB-regressor\textunderscore mixed}, that use the multiple target XGB regression for the state estimation architecture given in Fig. \ref{fig:xgbmultiple}. The former uses the training and validation sets of $X_{\textrm{noiseless\textunderscore pure}}$ as the input, and $Y_{\textrm{noiseless\textunderscore pure}}$ as the output. The latter uses those of $X_{\textrm{noiseless\textunderscore mixed}}$ and $Y_{\textrm{noiseless\textunderscore mixed}}$. The native Python API of XGB does not support multiple target regression, but it has features like \textit{DMatrix} data format for parallelization and computational speed up, automatic early stopping, and XGB's inherent cross-validation tool set. Therefore, instead of using XGB's Scikit-learn API and Scikit-learn's wrapper for multiple target regression, \textit{MultipleOutputRegression}, to combine single XGB regression models from the start at the training and model-selection (i.e hyper-parameter tuning) stages, we choose to train and tune 16 different models in XGB's native Python API, then combine them with \textit{MultipleOutputRegression} at the prediction (testing) stage.

We start the hyper-parameter tuning by a grid search of parameters using 5-fold cross-validation (CV) on the training set \cite{james2013introduction}. We split the training set into five subsets, and for five turns, we holdout one of these subsets for evaluation while training on rest and take the mean of these subset evaluations at the end. We, in the beginning, cross-validate while fitting a model only on the $\tau_0$ using 36 tomography measurements through choosing the \textit{objective} parameter in native Python API as \textit{reg:linear} for regression. We first create a 2D grid of parameters \textit{max\textunderscore depth} and \textit{min\textunderscore child \textunderscore weight} ranging from $9$ to $12$ and $5$ to $8$. Scanning through this grid while keeping learning rate $\eta$ = $0.3$, early stopping rounds of $10$, and everything else at default. We obtain the least root-mean-squared-error (RMSE, $\sqrt{\textrm{MSE}}$) scores as the evaluation metric. We find the best improvement in CV scores for a \textit{max\textunderscore depth} of $9$ and \textit{min\textunderscore child \textunderscore weight} of 6 with a RMSE of $0.099$ ($9.8\times10^{-3}$ MSE). The former parameter determines the maximum depth of XGB regression trees that are allowed to fit, and the latter determines the minimum sum of weights (Hessians) needed per `child' (when we do a split). Next, instead of optimizing all the hyper-parameters together, we save these results from the first 2D grid search and use them in our upcoming grid search.

This time we create another 2D grid ranging from $0.7$ to $1$ for the both dimensions to scan the \textit{subsample} and \textit{colsample\textunderscore bytree} parameters. The first parameter is to control what ratio of training instances (rows) are randomly sampled to grow trees (e.g., 0.5 uses half) to control over-fitting, and the latter does the same thing, but for columns. We found the ideal value for both of them to be $1.0$, leading to a CV score of $3\times10^{-2}$ RMSE ($9\times 10^{-4}$ MSE). Last, we search for optimal Ridge ($l_2$) and Lasso ($l_1$) penalty strengths, $\lambda$ and $\alpha$, by scanning through a square grid with both sides ranging from $0$ to $1$. The former regularization has the effect of shrinking the values of predictors (features) to a small but nonzero value, while the latter leads to parsimony (sparsity) due to completely turning of weights for some features. We found that $\lambda=0$ and $\alpha=0.4$ leads to a CV RMSE of $9\times 10^{-3}$ ($8.1\times10^{-5}$ MSE). Finally, using the previously optimized hyper-parameters, by executing a 1D grid search for learning rate $\eta$ ranging from $0.01$ to $0.3$, we found $0.1$ to be optimal with CV RMSE of $7\times 10^{-3}$ ($4.9\times10^{-5}$ MSE). We found no improvement by running the same model selection process through a 5-fold CV for the other 15 $\tau$ elements.

Using the optimal parameters found above, we fit 32 XGB regression models, 16 for noiseless pure states, and 16 noiseless mixed states. We set the number of rounds (a.k.a number of boosts, the equivalent of epochs for XGB) to 2000. However, due to early stopping, when validation set RMSE does not improve for ten rounds, training halts. We found most of the 32 models achieved very high accuracy (validation set MSE of the order $\sim 10^{-6}$), even more significant than neural nets, in less than 50 rounds (a couple of seconds in a laptop with a GPU). Some $\tau$ elements were harder to fit, and it took them around 200 rounds to fit with a validation set MSE of the order $\sim 10^{-3}$. Having trained 16 models for pure and 16 for mixed states, we combine their "boosters" (XGB models) within the Sci-kit learn \textit{MultipleOutputRegression} wrapper. Both the model for noiseless pure states and the mixed states achieve training, validation, and test set MSEs on the order $\sim 10^{-4}$, and an average fidelity $1.0$ with standard deviation $0.0$.

  \begin{center}
  \centering
 \begin{tabular}{||c c c c c||} 
 \hline
 \; & \textrm{Train MSE} & \textrm{Validation MSE} & \textrm{Validation Fidelity} & \textrm{Test Fidelity}\\ [0.5ex] 
 \hline\hline
 \textit{XGB-regressor\textunderscore pure}  & $9\times10^{-5}$ & $8\times10^{-5}$ & 1.0 & 1.0 \\ 
 \hline
  \textit{XGB-regressor\textunderscore mixed}  & $6\times10^{-5}$ & $5\times10^{-5}$ & 1.0& 1.0 \\ [1ex] 
 \hline
\end{tabular}
\label{tab:xgb_regress_tab}
\end{center}

The computational speed needed to fine-tune the model for extreme accuracy via reducing feedback time, processing data in a completely different way compared to linear models and neural networks, and ability to work with missing values and noise makes XGB a perfect candidate to work alongside and fix the mistakes of these other algorithms. In the QST context, there is an incentive to use a meta-model of different types of models, each with their advantage on interpreting tomography measurements, to tackle both the physical noise related to the measurement basis alignment and the statistical noise induced by the imputation procedure.

\ss{Meta-Model}

Ensemble learning is the method of using an ensemble of machine learning algorithms as a committee to make predictions \cite{Brown2010}. The boosting method we utilized is an ensemble learning method using subsequent models, where the latter models are trained on the former's output, though ensemble methods are not restricted to such dependent models. Stacked generalization is an ensemble learning approach that uses machine learning models that are trained using a training set as \textit{level-0} generalizers, and learn a \textit{level-1} meta-model on a hold-out (validation) data set to combine the prediction of the low-level models for predictive performance improvements that originators of the algorithm deemed `black art' \cite{wolpert1992stacked}. Low-level model predictions could be averaged or pooled as a weighted sum by giving models with higher accuracy higher weights, or their combinations for pooling can be learned by the meta-learner altogether \cite{ting1999issues}. Stacked generalization could use the same or different models as the \textit{level-0} base-learners, including neural networks, to boost the performance \cite{zhou2002ensembling}. Due to the resulting performance increase, researchers using stacking techniques have won awards in machine learning competitions such as The Netflix Prize \cite{bell2007bellkor}. In our case, we found that pure and mixed states can not be regressed together for state estimation, but they instead need their own set of machine learning models. We discovered that Conv1D, Conv2D, and XGB models have their own strengths and weaknesses while processing noisy and incomplete data due to the way they interact with features. Thus, we need one meta-learner of stacked base models for both pure and mixed states. This way, these stacked \textit{level-0} state estimation models cover each others' mistakes when the \textit{level-1} meta-model pools them.

We train two linear regression models to find the pooling weights for pure and mixed state models. In order not to over-fit, stacked generalization training is done using the validation set predictions \cite{wolpert1992stacked}. Concatenating a 50K validation set of noiseless pure states with 180K noisy states to get 230K measurements, resulting in 230K $\tau$ predictions from \textit{Conv1D-regressor\textunderscore pure}, \textit{Conv2d-regressor\textunderscore pure} and \textit{XGB-regressor\textunderscore pure} models and concatenating them, we obtain a $230K\times 48$ input matrix. We use this input matrix as the design matrix for the linear model. As the dependent target variables, we use the ground-truth $230K\times 16$  $\tau$ matrix elements. Fitting them for pure states, we obtain meta-model pooling weights $w_1=w_2=0.4$ and $w_3=0.2$. This means the \textit{Conv1D} and \textit{Conv2D} models generalize better for the noisy pure states than \textit{XGB}, yet all these models still cover each others' faulty predictions for better generalization performance. On the other hand, when we fit a linear meta-model for the mixed states, we found the pooling weights to be $w_1=w_2=0.16$ while $w_3=0.67$, meaning the \textit{XGB} model generalizes better for the mixed states.

\section{Results}

\begin{figure}[h!]
    \centering
    \begin{subfigure}{0.8\textwidth}
    \centering
    \includegraphics[width=\textwidth]{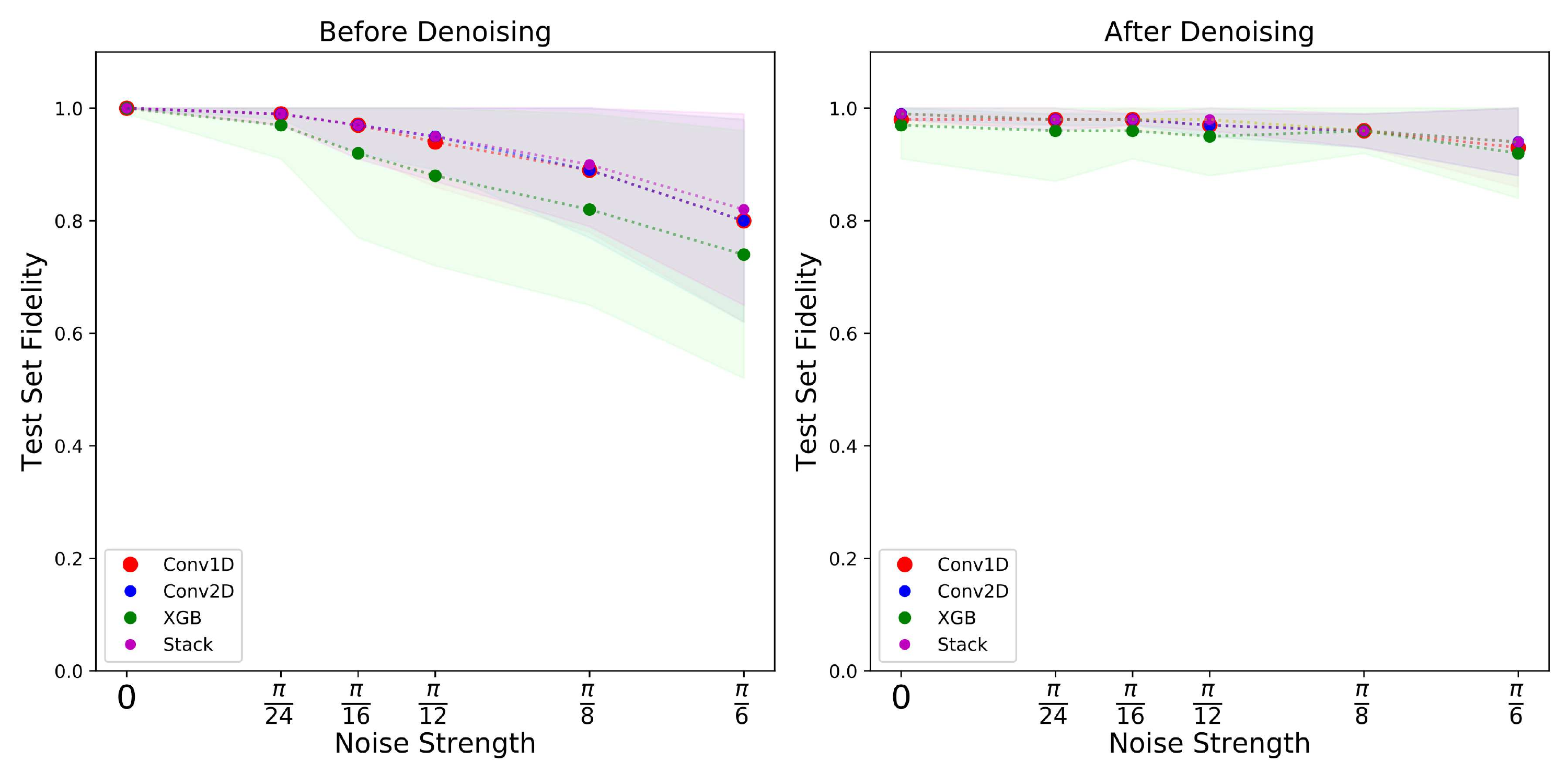}
    \caption{Pure states}
   \end{subfigure}
    \begin{subfigure}{0.8\textwidth}
    \centering
    \includegraphics[width=\textwidth]{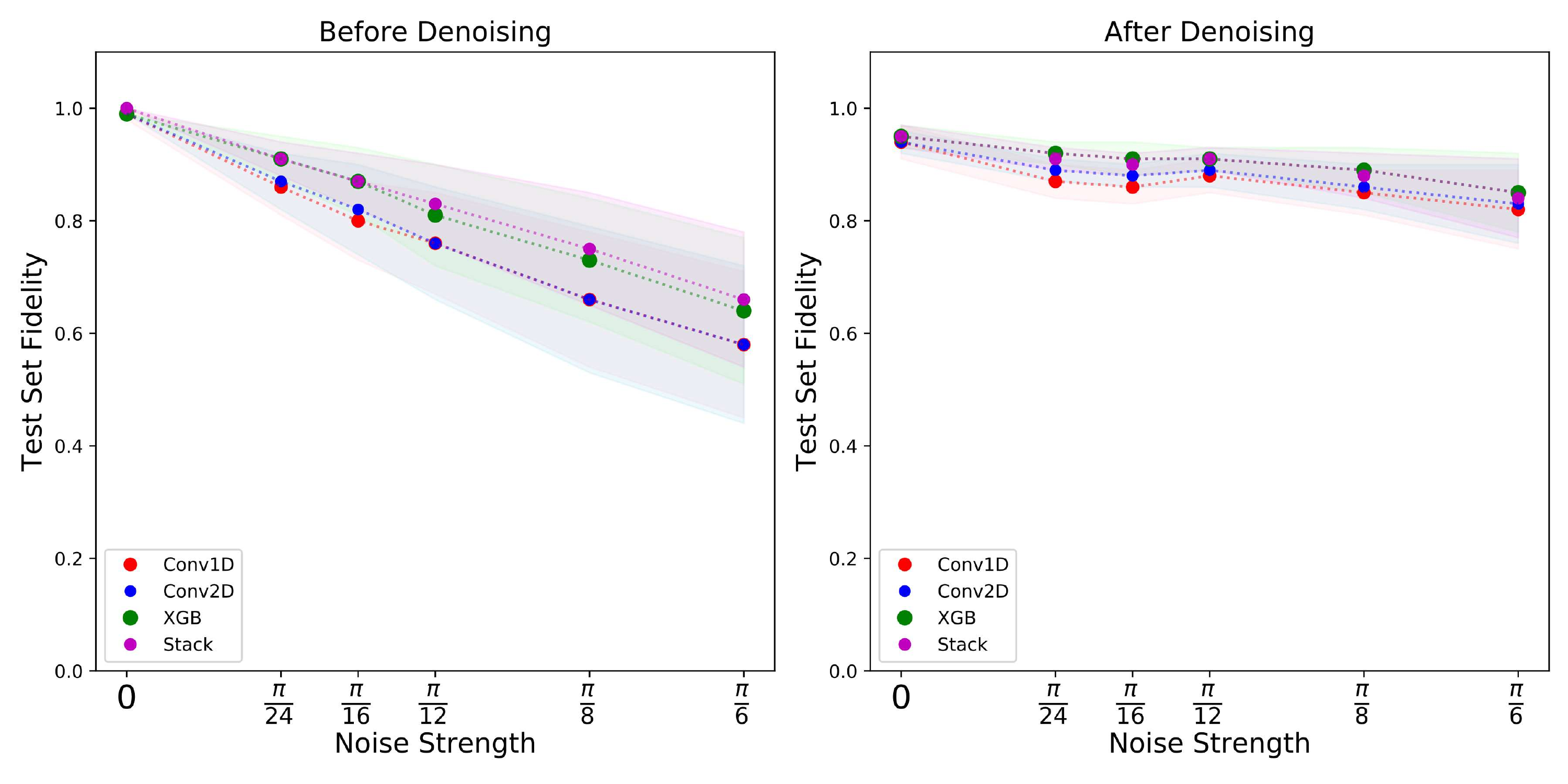}
    \caption{Mixed states}
   \end{subfigure}
    \caption{Test set fidelity vs. noise strength for a) pure and b) mixed states before and after denoising. Solid dots denote the average fidelity of the test set, while the colored hues denote the variance (standard deviation) of the fidelity. CNNs perform better for pure states, while XGB performs better for mixed states. Stacking and denoising increases the performance in all cases. Denoising when no noise is present decreases prediction accuracy.}
    \label{fig:noise_vs_fidelity}
\end{figure}

We start by assessing the performance of state estimation models with respect to noise with or without denoising. We generate one set of 10K noiseless measurements and five sets of 10K noisy measurements of noise strengths $\pi/24$, $\pi/16$, $\pi/12$, $\pi/8$, $\pi/6$ for pure states, and the same for mixed states. Plugging them into our QST models for pure and mixed states and their respective stacks, we, as expected, observed a decrease in predictive performance with the increasing noise strength, as shown in Fig. \ref{fig:noise_vs_fidelity}. We found that CNN based models are more resilient to noise for pure states, while the XGB based model has higher predictive power for mixed states, and the stacking of respective pure and mixed state models helps both. We observed in Fig. \ref{fig:noise_vs_fidelity} that with the increase of noise, not only does the average fidelity decrease (solid dots), but also the variance on fidelity (colored hues) increases, meaning noise on some states are tolerated even less. However, turning on the denoiser has a stabilizer effect. It both increases the average fidelities that were reduced with increasing noise and shrank the test set fidelity variances (hues in the figure). We also observe that denoising when no noise is present actually decreases performance; hence it underlines the importance of the detection of which measurement samples are noisy (\textit{Conv1D-isnoise}).

Next, we use Monte Carlo simulations via Halton sequences, explained in previous sections, to estimate the QST model performances when $k$ measurements are missing and recovered using \textit{MICE-BR}. We again create random masks to emulate missing measurements (50 masks for a single measurement missing, 100 for double that, etc.), calculate average test set fidelity for the measurements recovered by \textit{MICE-BR} after they were turned off for each mask, and compute the Monte Carlo mean of the average fidelity, and the MC standard deviation of the average fidelity.

\begin{figure}[t!]
    \centering
    \begin{subfigure}{0.8\textwidth}
    \centering
    \includegraphics[width=\textwidth]{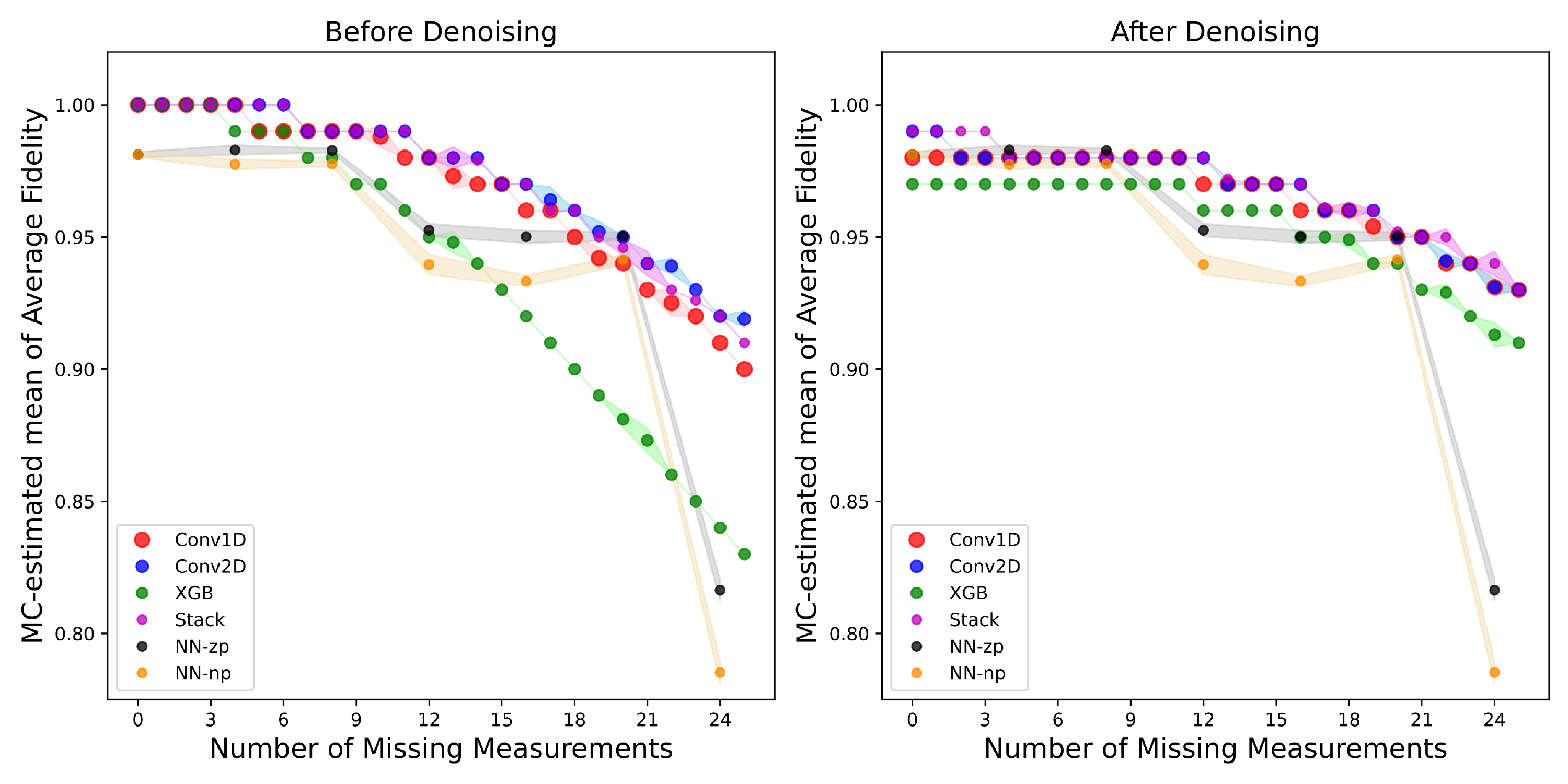}
    \caption{Pure states}
   \end{subfigure}
    \begin{subfigure}{0.8\textwidth}
    \centering
    \includegraphics[width=\textwidth]{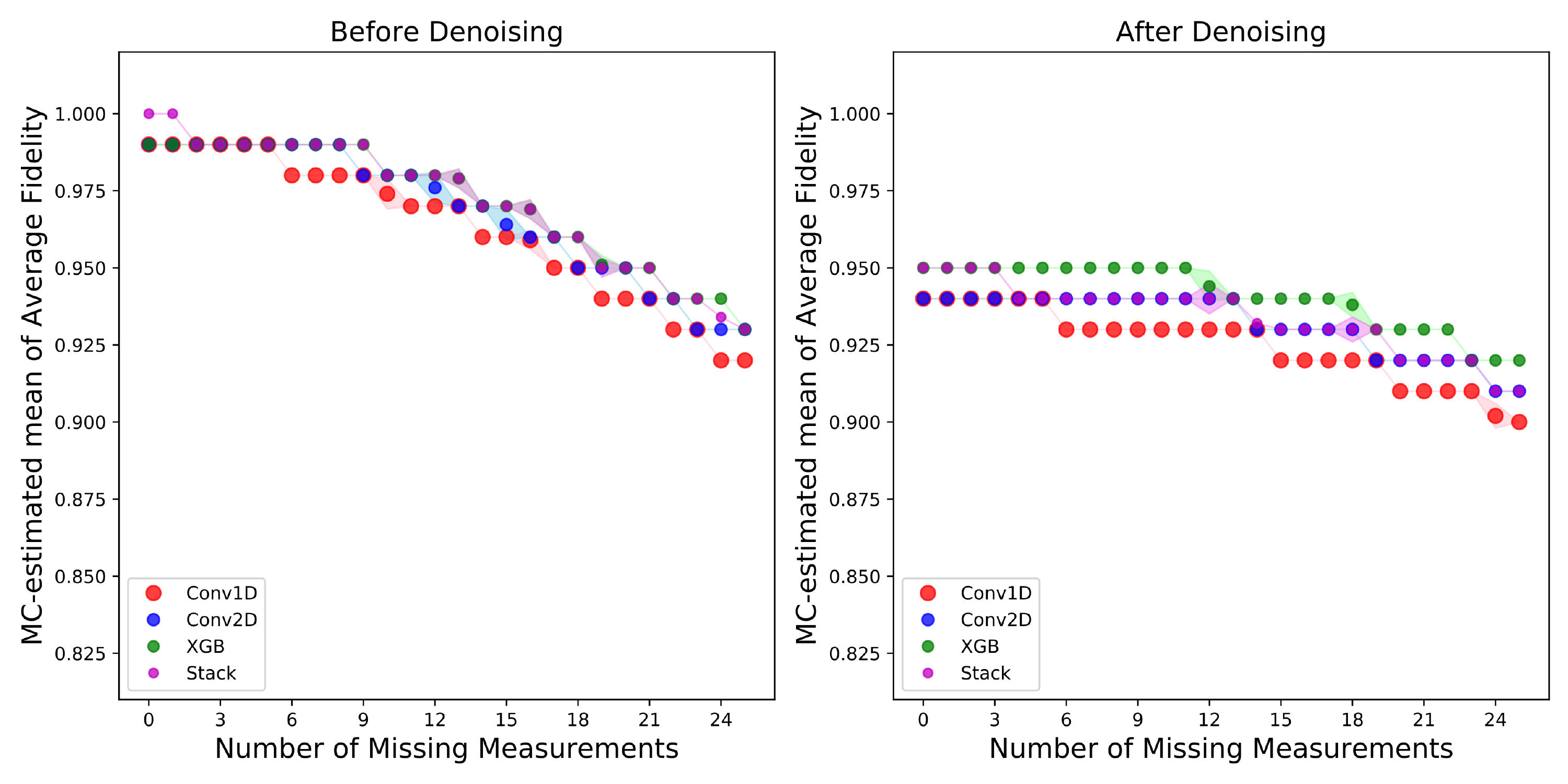}
    \caption{Mixed states}
   \end{subfigure}
    \caption{MC estimation for the estimation of noiseless a) Pure and b) Mixed states after $k$ measurements went missing on test set. Solid dots denote MC estimate mean for the average fidelity, while the hue denotes the variation of this average fidelity for different masks. We found stacking of models helps for both pure and mixed states. Passing a sample with imputed missing measurements through a denoising AE, which is originally trained using simulated experimental noise from basis rotations not the statistical imputation noise, before QST models in a pipeline improves reconstruction performance. Here "NN-zp" and "NN-np" denote the neural networks with zero-padding and no-padding results on pure states from our previous work for comparison \cite{lohani2020machine}. }
    \label{fig:noiseless_MC_fidel}
\end{figure}

\begin{figure}[h!]
    \centering
    \begin{subfigure}{0.8\textwidth}
    \centering
    \includegraphics[width=\textwidth]{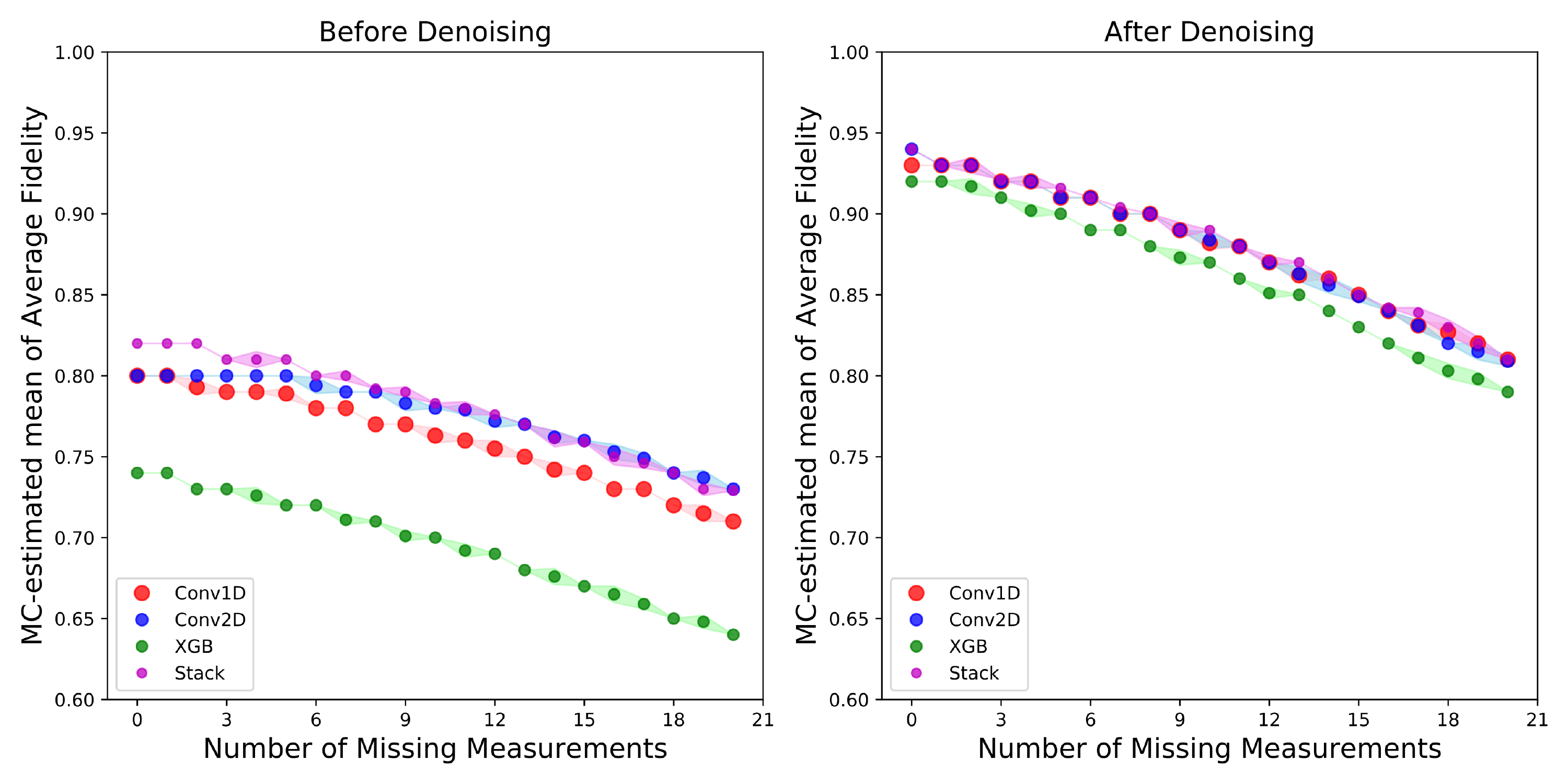}
    \caption{Pure states}
   \end{subfigure}
    \begin{subfigure}{0.8\textwidth}
    \centering
    \includegraphics[width=\textwidth]{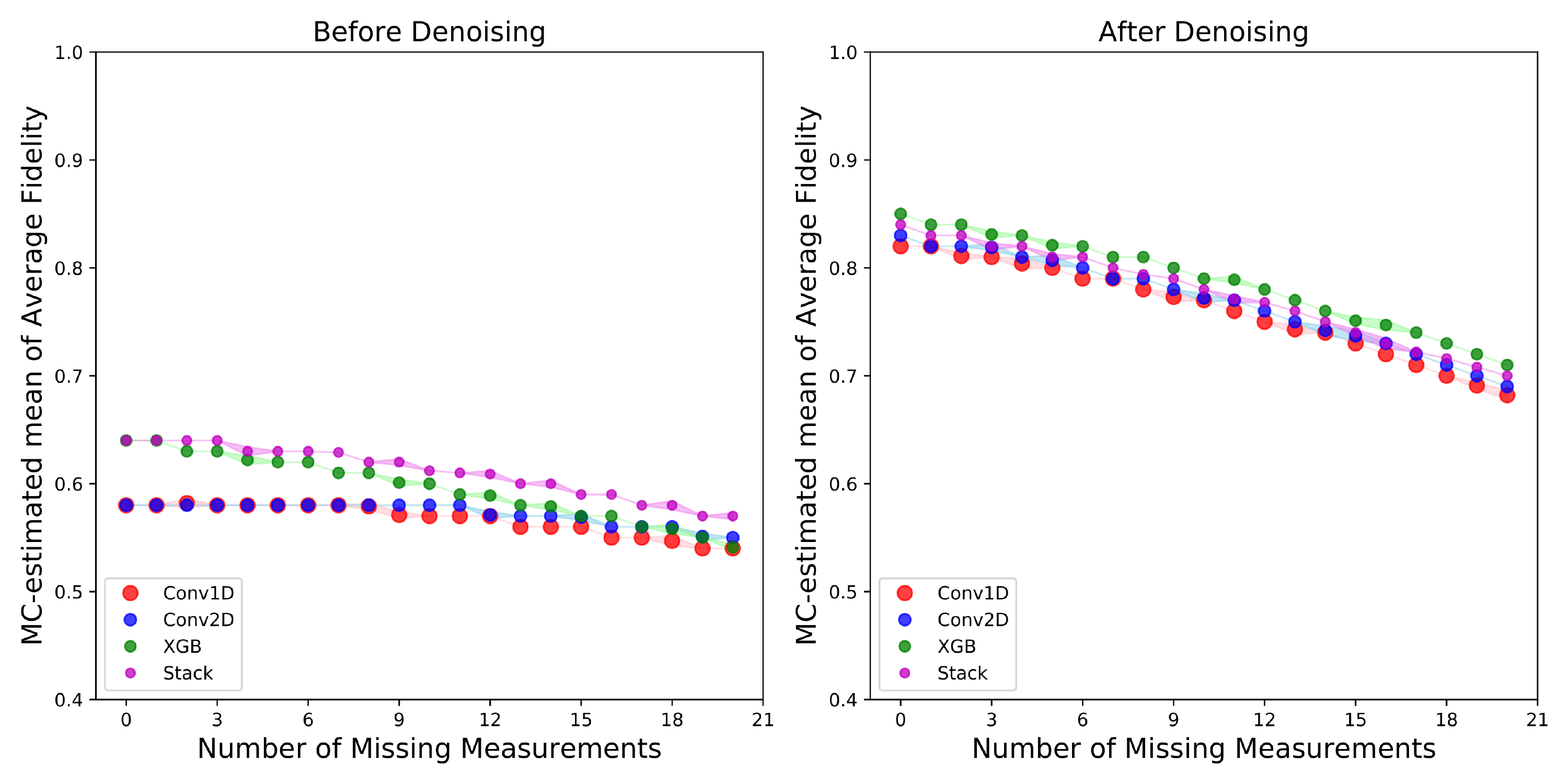}
    \caption{Mixed states}
   \end{subfigure}
    \caption{MC estimation for the estimation of a) Pure and b) Mixed states after imputation using noisy ($\pi/6$) measurements. Solid dots denote MC estimate mean for the average fidelity, while the hue denotes the variation of this average fidelity for different masks. We found stacking of models helps for both pure and mixed states. Passing a sample that contains both simulated experimental noise from basis rotations and imputed missing measurements through a denoising AE, which is originally trained using fiber noise not the statistical imputation noise, before QST models in a pipeline improves reconstruction performance.}
    \label{fig:noisepi6_MC_fidel}
\end{figure}

First, we use the MC average fidelity estimation scheme for the generated test sets of noiseless pure and mixed state measurements, see Fig. \ref{fig:noiseless_MC_fidel}. We found that the denoiser, which was trained using data coming from physical noise due to measurement basis misalignment during detection, not the MICE statistical noise, does not seem to improve imputation noise in terms of MSE between imputed and the original data (thus not included in that figure). Despite that, we found  \textit{Conv1D-denoise} actually improves prediction accuracy without improving MSE between $X_{\textrm{original}}$ and $X_{\textrm{recovered}}$. Even when $k=26$ measurements are missing (using only ten measurements), we still recover average fidelity values above $90\%$. We found that our current approach that combines imputation, stacking and denoising performs better than our previous approach that combines zero-padding or no-padding with 2D convolutional neural networks (Fig. \ref{fig:noiseless_MC_fidel}) \cite{lohani2020machine}. In our previous work we had two neural networks per \textit{k} missing measurements (e.g, for $k=1$ we replaced 36th measurement with zero for zero-padding, and used 35 dimensional input vectors for no-padding), our current approach not only has the advantage of compensating for $36\choose k$ combination of those missing measurements, but it also outperforms the previous models in terms of fidelity, especially when more than 20 measurements are missing. When we turn on the physical noise for these imputed measurements, the effects of stacking and the denoising AE are more pronounced for both the pure and the mixed states, as seen in Fig. \ref{fig:noisepi6_MC_fidel}.

\FloatBarrier



\section{Discussion}

In conclusion, we built a pipeline of machine learning models for quantum state estimation using projective measurements. Projective measurements coming to the pipeline have their missing values imputed, directed to denoising if they are detected to be noisy, and their quantum states estimated using different models depending on whether they are deemed to be pure or mixed states.

Unlike our previous work, we decoupled the handling of noise and the treatment of missing measurements from the models' training. This approach allows us to generalize our previous results to the case of any number and combinations of missing, noisy measurements, rather than the limited special cases.

\bibliographystyle{unsrt}
\bibliography{density}

\section*{Acknowledgements}
This material is based upon work supported by, or in part by, the Army Research Laboratory and the Army Research Office under contract/grant numbers W911NF-19-2-0087 and W911NF-20-2-0168. The views and conclusions contained in this document are those of the authors and should not be interpreted as representing the official policies, either expressed or implied, of the Army Research Laboratory or the U.S. Government. The U.S. Government is authorized to reproduce and distribute reprints for Government purposes notwithstanding any copyright notation herein.

\section*{Author contributions statement}
O.D. developed the neural networks and ran all simulations.  R.T.G. and B.T.K. conceived of and led the project.  O.D., R.T.G., S.L. and B.T.K. wrote the manuscript. All authors contributed to the discussions and interpretations of the results.

\section*{Data availability}
The data that support the findings of this study are available from the corresponding authors on reasonable request.

\section*{Competing interests}
The authors declare no competing interests.

\end{document}